# Growth and electronic properties of nanolines on TiO$_2$-terminated SrTiO$_3$(001) surfaces

W. Yan,[1] W. Sitaputra,[2] M. Skowronski[1] and R. M. Feenstra,[2,a]

[1]*Materials Science and Engineering Department, Carnegie Mellon University, Pittsburgh, PA, 15213, USA*

[2]*Department of Physics, Carnegie Mellon University, Pittsburgh, PA, 15213, USA*

[a]Electronic mail: feenstra@cmu.edu

Surfaces of homoepitaxially grown TiO$_2$-terminated SrTiO$_3$(001) were studied *in situ* with scanning tunneling microscopy and spectroscopy. By controlling the Ti/Sr ratio, two-dimensional domains of highly ordered linear nanostructures, so-called "nanolines", are found to form on the surface. To further study how the surface structure affects the band structure, spectroscopic studies of these surfaces were performed. Our results reveal significantly more band bending for surfaces with the nanolines, indicative of an acceptor state associated with these features. Additionally, an in-gap state is observed on nanoline surfaces grown under high oxygen deficient conditions. This state appears to be the same as that observed previously, arising from the (++/+) transition level of surface oxygen vacancies.

## I. INTRODUCTION

As a common complex-oxide substrate, SrTiO$_3$ has been intensely investigated for several decades. Since 2004 in particular, the SrTiO$_3$ surface has drawn particular attention from researchers owing to a discovery of a 2D electron gas (2DEG) at the interface of a complex-oxide heterostructures, i.e. LaAlO$_3$(001)/SrTiO$_3$(001).[1-6] In such heterostructures, where novel properties are found at the interface, it is important to be able to distinguish any contribution from the substrate, SrTiO$_3$ (STO), from those that are produced by formation of the epitaxial overlayer, LaAlO$_3$ (LAO). Therefore, it is crucial to have a complete material concerning the surface of the substrate immediately prior to the overlayer growth. Castell *et al.* demonstrated that by argon-ion sputtering and ultra-high-vacuum (UHV) annealing[7,8,9] (as well as by Ti deposition and subsequent oxidation),[10] the STO surface will form close-packed domains of linear features, denoted as *nanolines*. These features were shown to be composed of TiO$_2$-derived complexes, and they thus provide new means for developing technologies for photocatalysis.[9] Those workers studied the electronic properties of the nanolines by x-ray photoelectron spectroscopy and density functional theory,[9] but a study of such properties by scanning tunneling spectroscopy (STS) has not been performed to date.

It is well known that homoepitaxial growth by molecular beam epitaxy (MBE) can yield surfaces of high surface quality, and it enables fine control of their stoichiometry. In this work, we study MBE-grown TiO$_2$-terminated SrTiO$_3$(001) surfaces *in situ* using scanning tunneling microscopy (STM) and STS. With the combination of STM and MBE, effects on the surface structure arising from slight changes in stoichiometry can be readily detected, similar to the case of compound semiconductors.[11,12] We find that by increasing the Ti/Sr flux ratio during growth, highly ordered linear features were observed to form on $TiO_2$-terminated STO surfaces; these features are identical to the *nanolines* observed by Castell and co-workers.[7,8,9] The fine stoichiometric control enabled by MBE in our experiment permits the controllable formation of the nanoline structures. Using STS, we have studied the impact of the nanolines on the electronic properties of the surface. We find nanolines introduce acceptor states at the surface (negative when occupied by an electron), resulting in upwards band bending for n-type material. Additionally, on nanoline surfaces grown with high oxygen deficient conditions ($1 \times 10^{-6}$ Torr) an in-gap state is observed, which we associate with the transition level of surface oxygen vacancies that we identified in a prior study.[13] Our findings thus yield further insight on homoepitaxial surfaces of $SrTiO_3(001)$ and their unusual electronic behavior, which may be related to the question of whether or not such surfaces produce any effects on the 2DEG formed at the interface after the subsequent growth of other oxides.

## II. EXPERIMENTS

Substrates are 0.05 wt % Nb-doped $SrTiO_3(001)$ wafers, which were prepared using the Arkansas procedure to obtain a $TiO_2$-terminated surface with a well-defined step-terrace structure.[6] Homoepitaxial growth of the $SrTiO_3(001)$ was done at a temperature of 800 °C under two oxygen pressure conditions: $5 \times 10^{-6}$ Torr and $1 \times 10^{-6}$ Torr, using our home-built MBE system (see details in the Supplementary Material) which contains a leak valve for molecular oxygen, a Ti-ball for titanium, and an effusion cell for strontium. Following growth, the samples were transferred into the STM chamber without breaking vacuum. STM and STS measurements were performed at room temperature with a Pt/Ir tip. Each spectrum requires several seconds to acquire, and the results shown below consist typically of an average of four such spectra. The conductance, dI/dV, measured by a digital lock-in amplifier method, using a modulation frequency of typically 15017 Hz and rms (root-mean-square) amplitude of 32 mV. We use an STS measurement method that yields relatively high dynamic range, described in detail in Ref. [14], in which the tip-sample separation is varied as a function of applied sample bias voltage $V$. A ramp is applied to the tip-sample separation according to $\Delta S(V) = a \, (|V| - |V_0|)$, where the constant $a$ is typically taken to be 2 – 3 Å/V for our measurements and $V_0$ refers to the sample bias for the constant-current operation of the STM immediately preceding the acquisition of the spectrum. The exponential increase in conductance arising from this variation in tip-sample separation is then normalized by multiplying the data by a factor of $e^{2\kappa \Delta S(V)}$, where $\kappa$ is an experimentally determined decay constant. Values of $\kappa$, measured in each experiment, were found to be in the range of 4 – 6 nm$^{-1}$ (this value is somewhat less than an ideal value of about 10 nm$^{-1}$, for reasons discussed in our previous paper).[13] This normalization does not affect any detailed structure in the spectra, but it improves the dynamic range by several orders of magnitude, thus yielding enhanced detection sensitivity for band-edge features and/or in-gap states (use of the lock-in amplifier is essential for this method, since it yields the derivative of current with respect to voltage at *fixed* tip-sample separation, at each voltage, as discussed in Ref. [15]). A noise level for the conductance is measured with the tip withdrawn from the surface; subsequent normalization with the same method then yields a voltage-dependent noise level for each spectrum.

## III. RESULTS AND DISCUSSION

By performing growths under a range of Ti/Sr flux ratios, we have observed the evolution of the surface morphology from TiO$_2$-terminated flat surfaces to ones covered with nanolines, as shown in Fig. 1. The TiO$_2$-terminated flat surfaces are obtained by performing the growth with a Ti/Sr flux ratio near unity and stopping the growth at a maximum in the intensity reflection high-energy electron diffraction (RHEED) oscillations, as described in the Supplementary Material. Surfaces prepared in this way display either a c(4×2) or a (2×1) RHEED pattern (an example of the former is shown in Fig. S2), indicative of double-layer TiO$_2$ termination.[16,17,18] These surfaces are found to have flat terraces extending over ~100 nm, as seen in Fig. 1(a). The terraces are separated by full-unit-cell high (0.40 nm) steps as shown in Fig. 1(b), indicative of uniform TiO$_2$ termination (as opposed to the mixed termination found in our prior study of cleaved surfaces).[13,19] The step edges have a jagged morphology, extending along [100] and [010] directions, Fig. 1(a), consistent with prior reports.[16,20,21] For sufficiently high Ti-rich conditions (obtained by reducing the temperature of the Sr cell), the nanoline features are found to form, as shown in Figs. 1(c) and 1(d), very similar to those previously reported by Castell and co-workers[7,8,9] (Fig. 1(c) is the same data as shown in our prior work).[13] These features were found to form regardless of the amount of oxygen deficiency in the growth. An enlarged image in the blue frame in Fig. 1(d) is shown in Fig. 1(e). Linear corrugation rows with the width of approximately 3.2 nm and height of 0.2 nm (corresponding to half of the STO lattice constant) are found. The length of the nanolines exceeds 60 nm. Detailed atomic arrangements in the rows of Fig. 1(e) are not apparent in the image, due to electronic and mechanical noise in our instrument, so we cannot distinguish between the dilines, trilines, and tetralines reported Castell and co-workers.[7,8,9] Nevertheless, based on their width and corrugation amplitude, we believe the linear features we observe to be most consistent with the triline nanostructures. Our results clearly show that the surface morphology depends sensitively on the stoichiometry of the surface.

To investigate how the electronic properties of the surface changes with the surface structure, we first studied the c(4×2)/(2x1) $TiO_2$-terminated surfaces grown with oxygen partial pressure of $5 \times 10^{-6}$ Torr. As shown in Fig. 2(a) and 2(c), the surface morphology of the two samples are indicative of growth under slightly Ti-rich conditions, with large flat terraces are separated by jagged steps edges with step height of 0.40 nm. Figures 2(b) and 2(d) shows averaged dI/dV vs. V spectra acquired on the two surfaces, respectively. The spectra are quite reproducible, and they reveal that the Fermi level (corresponding to 0 V in the spectra) is located near the onset of a band extending up to positive voltage (i.e. empty states). The theoretical density-of-states for the c(4×2) surface reveals a band gap that is the same, within 0.1 or 0.2 eV, as the bulk gap,[9] and similarly for the closely related (2×1) surface.[22] In particular, for the latter surface the conduction band (CB) edge of that surface is reported to maintain its Ti character associated with bulk-like Ti atoms in the STO.[22] Hence, we can identify the band at positive voltages in Figs. 2(b) and 2(d) as arising from the CB edge of bulk STO. This result indicates essentially flat bands (i.e. no band bending) for these surfaces, which is in contrast to the Fermi level position of cleaved STO surfaces that is found to be pinned near the middle of the band gap, due to states arising from various types of surface disorder (and from oxygen vacancies, when present).[13,18]

STO samples growth by higher Ti/Sr flux ratios (relative to those that produce the c(4×2)/(2x1) $TiO_2$-terminated surfaces of Fig. 2) in oxygen partial pressure of $5 \times 10^{-6}$ Torr is found to produce the nanolines on the surface, as shown in Fig. 3. The Ti/Sr ratio increases from Fig. 3(a) to Fig. 3(d) (by decreasing the temperature of the Sr source, while maintaining the current of the Ti source at fixed value). Nanolines with widths of 2.9 – 3.4 nm (full width at half height), heights of 0.18 – 0.23 nm, and with various lengths are formed. STM images of the nanolines are bias dependent for low sample bias, as shown in Fig. S3 in the Supplementary Material. As discussed above in connection to Fig. 1, these linear features share a strong similarity both in terms of width and corrugation amplitude to the typical triline nanostructures, for which Castell and co-workers have provided STM images of exceptional clarity.[7,8,9] Those workers have argued that this reconstructed surface is composed of a several $TiO_x$ overlayers on top of $TiO_2$-terminated STO. This conclusion is consistent with our observation that the nanolines only begin to appear for Ti-rich growth conditions.

To study how the nanolines affect the electronic properties of the surface, averaged STS spectra are shown in Fig. 3(e)–(h). These spectra have a common feature that the Fermi level is pinned near midgap, with the onset of CB shifted upwards by around 1.5 eV compared with the c(4×2)/(2×1) surfaces of Fig. 2. The nanolines thus induced band bending of the surfaces, further discussed below. Since the samples started as n-type prior to the formation of nanolines (i.e. as demonstrated in Fig. 2), the observed band bending indicates that the nanolines introduce *acceptor states* on the surface (i.e. states that are negative when occupied by an electron), consistent with the results of Castell *et al.* From x-ray photoelectron spectroscopy (XPS), they detected a well-defined mid-gap state associated with the nanolines, at 1.0 eV below the Fermi energy (with the Fermi level near the CB minimum).[9]

Our STS results, Figs. 3(e)–3(h), do not directly reveal such an acceptor state. However, we consider it quite possible that the acceptor states are of insufficient density to permit observable current through them to be attained in the measurement, as found previously for other large-band-gap surfaces.[23,24] On the surface of STO, as just argued, acceptor states will be produced with the formation of nanolines. Such states accept electrons from the CB, producing upwards band bending in the semiconductor and a Fermi level position at the surface that is expected to lie within the acceptor band. For a sufficiently high concentration of the acceptor states, the carriers will be able to move among the states (and/or from bulk to surface), and one expects that the states will be directly observable, as seen e.g. for the donor states induced by a high density of oxygen vacancies on the STO surface and for other systems.[13] However, when transport between surface states is limited, then they may not be able to refill fast enough to supply current to the tip, in which case the acceptor band will not visible in the STS measurement. We note in passing that the main spectra of Fig. 3, as well as those in Fig. 2, do not reveal any signal from the valence band (VB); such a signal is expected to occur about 3.2 V (the STO band gap) below the CB onset, hence lying just at the edge, or outside, of the measurement range. However, as shown by the inset of Fig. 3(h), a strong signal that we associate with the VB edge can be seen when the measurements are extended to larger negative voltages.

Aside from the nanolines, we further found by STS that other factors like oxygen vacancies may change the electronic properties of STO dramatically. As shown in Fig. 4(a), clear contrast arising from nanolines is still found on STO samples grown using high Ti/Sr flux ratios and with high oxygen deficient conditions (1 × 10$^{-6}$ Torr). A peak located at about 1 eV below the CB edge is clearly observed, as shown in Fig 4(b). However, for spectra acquired on STO surfaces grown with higher oxygen partial pressure of 5 × 10$^{-6}$ Torr, as shown in Fig 4(d), in-gap states are not observed, which is consistent with the dI/dV curves in Fig 3. (For the spectra of Figs. 4(b) and (d), only the current was measured, not dI/dV, but nevertheless the interpretation of current vs. voltage is quite similar to that for dI/dV spectra). Based on these findings, we interpret the in-gap peak as arising from surface oxygen vacancies on the $TiO_2$-terminated surfaces.

The in-gap peak seen in Fig. 4(b) is very similar to what we have previously observed for oxygen vacancies on SrO-terminated STO surfaces.[13] In that prior work we did not, however, observe such states on $TiO_2$-terminated surfaces (even though those surface were studied and were expected to contain a relatively high number of vacancies). An important difference between the spectra acquired in the present experiment compared to the prior work is the magnitude of the ramp used in the tip-sample separation during acquisition of the spectra, 3 Å/V for Fig. 4 compared to typically 1.5 Å/V in our prior work.[13] Therefore, the sensitivity to in-gap states is increased in the present work, and we believe that this difference in measurement conditions is what enables us to now observe the oxygen vacancy peak for the $TiO_2$-terminated surfaces. Indeed, we demonstrated theoretically in our prior work that similar in-gap states are expected to be present on both SrO-terminated and $TiO_2$-terminated surfaces, but the wave functions of the latter type of states are confined much nearer to the surface than the former, which we argued was the reason that we could not observe them in that prior work.[13] The successful observation of the states on $TiO_2$-terminated surfaces in the present work, using smaller tip-sample separations, lends credence to that argument.

Based on the recent theoretical work of Janotti *et al.*,[25] we interpret the in-gap spectral peak as arising from a transition from the 2+ charged state of an oxygen vacancy to the + state which includes an electron in a nearby bound polaron state.[13,19] The dependence of the in-gap peak on oxygen growth pressure show that the number of oxygen vacancies can be controlled by the oxygen pressure during the growth. The STO surfaces grown with low oxygen partial pressure contain a high number of vacancies, while surfaces grown with high oxygen partial pressure contain a relatively low number of vacancies.[26,27] One important aspect of our observed in-gap peaks, both in the present work and in our prior work, is that they are consistently located *above* the Fermi level.[13] This position above the Fermi level might be taken as an indicator of acceptor-like nature (i.e. negative when filled with electrons) of the states, which would be in contradiction to the theoretical identification of donor-like behavior (i.e. positive when empty of electrons) of the vacancy states. Alternatively, as we have previously argued, if there are additional acceptor-like states located in the lower part of the gap, then electron transfer will occur from the vacancy-induced donor states to the lower lying acceptor states, resulting in the vacancy-induced donor states lying *above* the Fermi level. In our prior work we could not identify the nature of these low-lying acceptor states, although experimentally they were found to be more numerous on somewhat rough surfaces, so we referred to them generally as "disorder induced". In the present work, our surfaces are quite well ordered in the absence of nanolines, as seen in Fig. 2. Furthermore, as argued in connection with Fig. 3, the nanolines themselves induce acceptor-like states on the surface. Hence, electron transfer from the vacancies to the nanolines may well be occurring, resulting in the vacancy-induced states lying above the Fermi level (or, possibly, additional disorder-induced acceptor states may also be present on the surface). We note that the electron polarons associated with oxygen vacancies at $TiO_2$ rutile and anatase surfaces observed by Setvin *et al.*[28,29] were found to have in-gap states lying at 0.7 eV and 1.0 eV, respectively, *below* the Fermi level. Those surface appears to be of relatively high quality, i.e. without significant surface disorder, and hence the vacancy-induced states lie at the theoretically expected location below the Fermi level.

## IV. CONCLUSION

In summary, by increasing the Ti/Sr ratio during epitaxy growth, highly ordered linear features denoted as nanolines can be formed on Ti-rich $TiO_2$ terminated surfaces. Such features are found to lead to Fermi level pinning near the middle of the band gap (believe to arise because of acceptor-type defects of the nanolines,[9] although such defects are not directly revealed in our STS measurements). Additionally, an in-gap transition level contributed to oxygen vacancies is observed on nanoline surfaces grown with low oxygen partial pressure. Overall, our MBE-grown surfaces provide an effective method for control of the nanoline structures, hence providing control of the electronic properties of the surface.

## SUPPLEMENTARY MATERIAL

See supplementary material for experimental details and bias-dependent images of the nanolines.

## ACKNOWLEDGMENTS

The authors thank Varun Mohan for his technical assistance in this work. This research was supported by AFOSR Grant No. FA9550-12-1-0479.

**FIGURES**

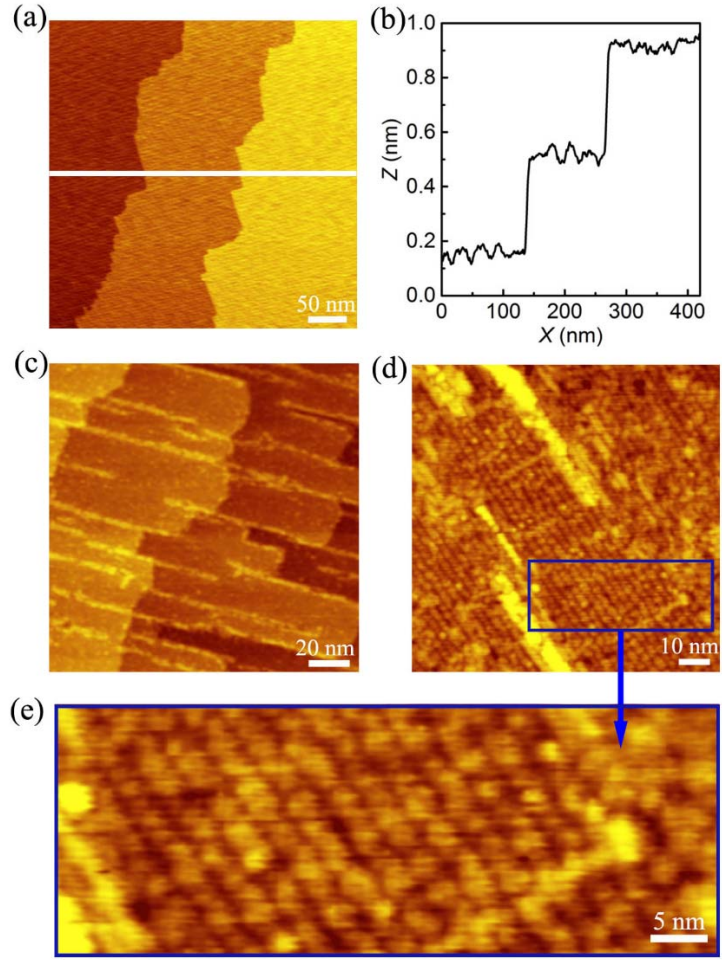

FIG. 1. STM images illustrating different topographic features observed on SrTiO$_3$ surface for different surface stoichiometries: (a) c(4×2)/(2×1) surface, TiO$_2$ terminated, (b) surface profile along line indicated in (a), (c) –(e) surfaces grown under Ti-rich conditions. Panel (e) shows an expanded view of the area indicated in (d). Acquisition parameters (sample voltage and tunnel current) for the data are: (a) and (b) $V_s = 2.1$ V , $I = 0.54$ nA; (c) $V_s = 1.5$ V , $I = 0.39$ nA; (d) and (e) $V_s = 3.1$ V , $I = 0.03$ nA. Images are displayed with surface height denoted by the color scales, over ranges of (a) 10, (c) 16, (d) 7, and (e) 4 Å.

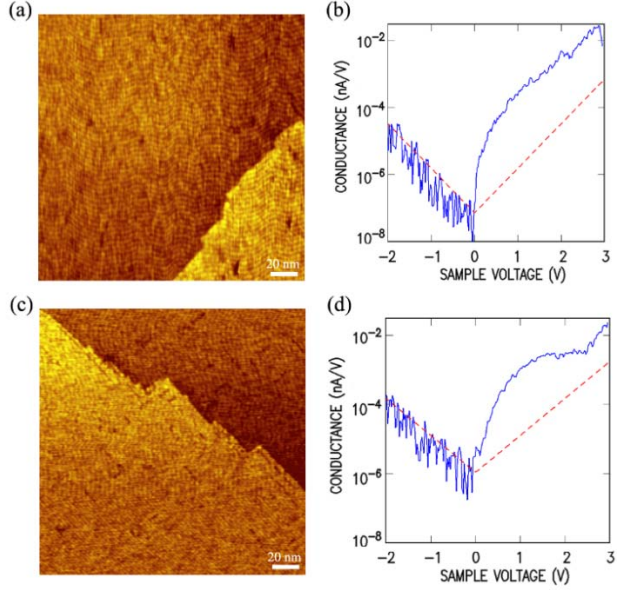

FIG. 2. (a) and (c) 200 × 200 nm$^2$ topography images of c(4×2)/(2×1) TiO$_2$-terminated SrTiO$_3$ surfaces grown in oxygen partial pressure of 5 × 10$^{-6}$ Torr. Acquisition parameters for the data are: (a) $V_s$ = 2.9 V, $I$ = 0.04 nA, (c) $V_s$ = 2.9 V, $I$ = 0.04 nA, and the image are displayed with height-ranges of 7 and 6 Å, respectively. (c) and (d) are averaged dI/dV-V spectra acquired on (a) and (c), respectively. The red dotted lines indicate the voltage-dependent noise level for each measurement.

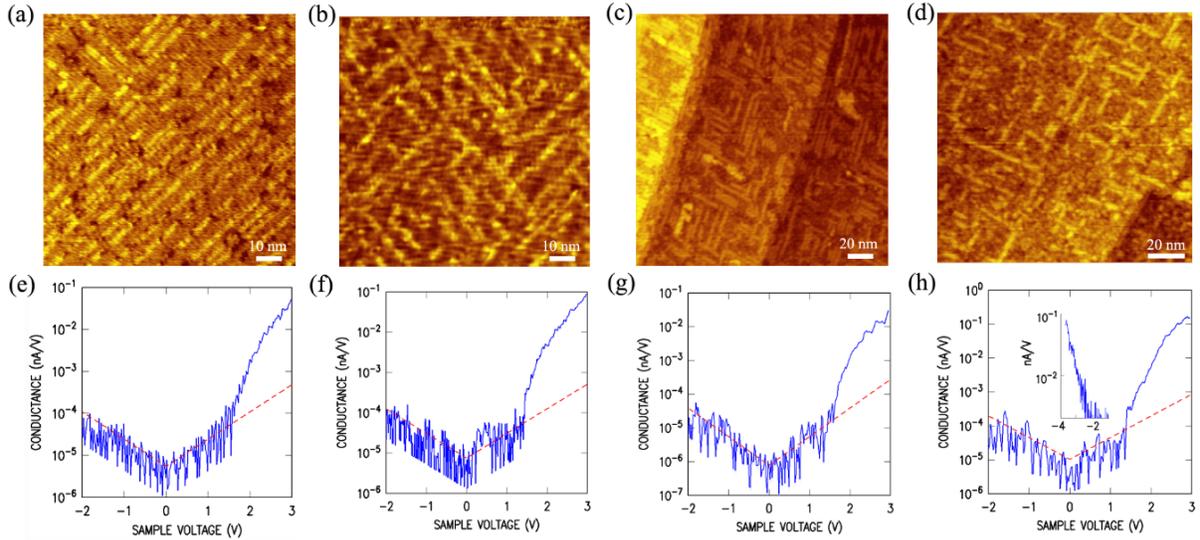

FIG. 3. (a)–(d) STM images of four Ti-rich TiO$_2$-terminated STO surfaces grown with oxygen partial pressure of 5 × 10$^{-6}$ Torr. The Ti/Sr ratio increases from (a) to (d) controlled by decreasing the temperature of Sr cell while maintaining the current of the Ti cell at a fixed value. Images sizes are 100 × 100 nm$^2$, 100 × 100 nm$^2$, 200 × 200 nm$^2$, 138 × 138 nm$^2$ for (a)–(d), respectively. Acquisition parameters for the data are: (a) $V_s$ = 2.7 V, $I$ = 0.02 nA, (b) $V_s$ = 2.8 V, $I$ = 0.03 nA, (c) $V_s$ = 2.9 V, $I$ = 0.05 nA, (d) $V_s$ = 3.1 V, $I$ = 0.04 nA, and the image are displayed with height-ranges of 3, 4, 11 and 7 Å, respectively. (e)–(h) Averaged spectra acquired on (a)–(d) respectively. The red dotted lines indicate the voltage-dependent noise level for each measurement. The inset of (h) shows a spectrum acquired with constant tip-sample separation (i.e. no ramp) and extending out to larger negative sample bias, showing the VB edge.

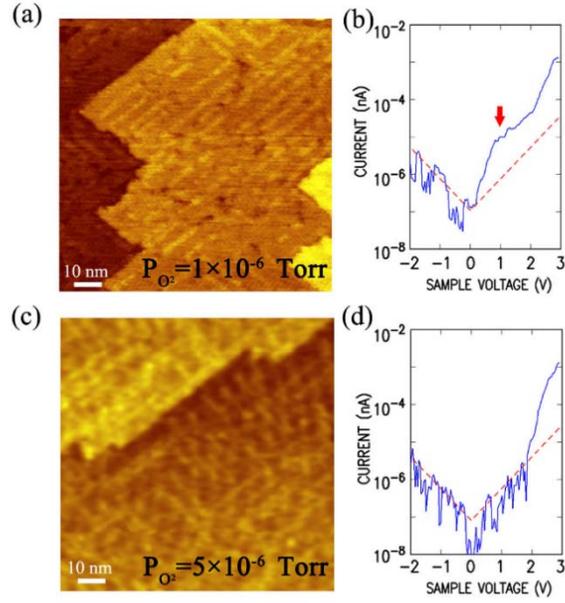

FIG. 4. STM images illustrating nanoline contrast for homoepitaxial surfaces of SrTiO$_3$(001) grown with oxygen partial pressure of (a) $1 \times 10^{-6}$ Torr and (c) $5 \times 10^{-6}$ Torr. The images were acquired with a tunneling current of 0.03 nA and sample bias voltage of 3.0 V. Image size is $100 \times 100$ nm$^2$ for both images. The corresponding height-ranges are 9 and 11 Å respectively. (b) Averaged current vs voltage spectra acquired on surface of (a), showing an in-gap peak (as indicated by the red arrow). (d) Averaged current vs voltage spectra acquired on surface of (c). Red dotted lines indicate the voltage-dependent noise level for each measurement.

# Supplementary material

# Growth and electronic properties of nanolines on TiO$_2$-terminated SrTiO3(001) surfaces

W. Yan,[1] W. Sitaputra,[2] and M. Skowronski[1] R. M. Feenstra[2]

[1]Materials Science and Engineering Department, Carnegie Mellon University, Pittsburgh, PA, 15213, USA

[2]Department of Physics, Carnegie Mellon University, Pittsburgh, PA, 15213, USA

**Experimental details:**

0.05 wt % Nb-doped SrTiO$_3$(001) wafers are bought from CrysTec. The substrate was etched in Aqua Regia solution, and then annealed *ex situ* for 1 hour at 1000 ˚C in air. By using this Arkansas procedure, SrO on the surface is removed and a TiO$_2$-terminated surface is obtained [1]. Homoepitaxial growth of the SrTiO$_3$(001) was done in our home-built molecular beam epitaxy system. Sources consisted of a leak valve for molecular oxygen, a Ti-ball for titanium and an effusion cell for strontium. The Sr and Ti were co-deposited under two different oxygen partial pressure ($1 \times 10^{-6}$ Torr and $5 \times 10^{-6}$ Torr). Using the methods described below, flux ratios between Sr and Ti were maintained at unity or slightly on the Ti-rich side, thus avoiding Sr-rich conditions which necessarily lead to SrO termination of the surface as discussed by Nie *et al.* [1].

Initial calibration of our Sr and Ti flux ratio was performed using x-ray diffraction (XRD) of 50- or 100-monolayer-thick films, as shown in Fig. S1. Satellite peaks arise in the XRD spectra when growth is performed under significantly Sr- or Ti-rich conditions, due to lattice variation caused by excess cations, as shown in Figs. S1(a) and S1(c). In contrast, for Fig. S1(b) no satellite peaks are found, and we identify this condition as corresponding to stoichiometric growth.

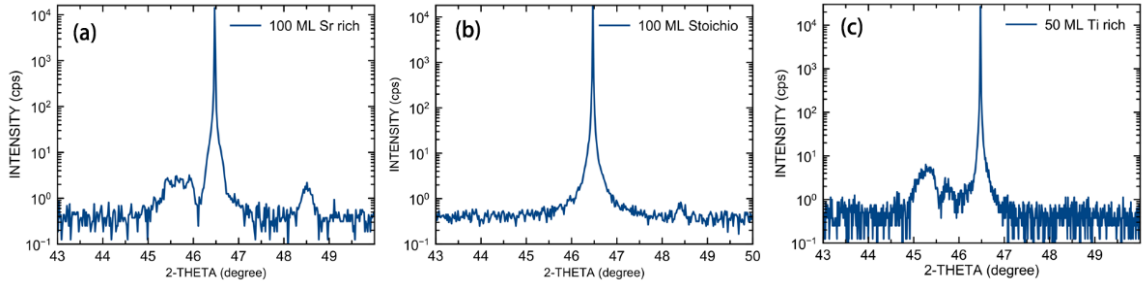

FIG S1. XRD spectra for different growth conditions of SrTiO$_3$(001) homoepitaxy. Films are grown with fixed Ti flux and with Sr cell temperatures of (a) 475, (b), 465, and (z) 455 °C.

For sample prepared for our study of nanolines, typically 5 monolayers of SrTiO$_3$ was grown, using reflection high-energy electron diffraction (RHEED) patterns and intensity oscillations to monitor the growth in real time. As shown in Fig. S2(a), clear RHEED oscillations are observed, indicative of a layer-by-layer growth mode. Starting from a TiO$_2$-terminated substrate, the deposition starts when the shutter is opened. After each RHEED oscillation, one STO layer is formed on the substrate, and the surface returns to being TiO$_2$ terminated. Hence, when the shutter is closed at the maximum of an oscillation, such as the fifth oscillation in Fig. S2(a), then the resulting, exposed surface of STO is expected to be TiO$_2$ terminated. Confirmation of this termination is made by observation of the RHEED pattern; we observe either c(4×2) or (2×1) symmetry, with an example of the former shown in Fig. S2(b). These surface structures are closely related, with both of them having double-layer $TiO_2$ termination [3,4]. In contrast to the distinct oscillations of Fig. S2(a), growth with the Sr cell temperature 1 °C lower then yields the RHEED intensity curve of Fig. S2(c), in which the oscillations have greatly diminished in amplitude. In this case, when the growth is terminated, a faint high order reconstruction is observed by RHEED. Such films are found by STM to be covered in the Ti-rich nanolines. After the growth, all samples were cooled down to room temperature. Then, the samples were transferred *in situ* into the STM chamber for further measurement.

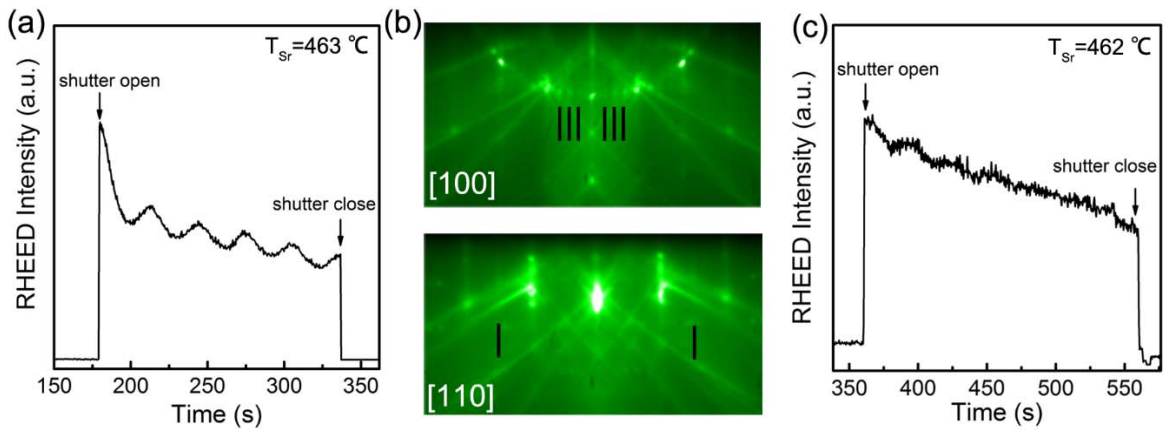

FIG S2. RHEED intensity oscillations of the specular spot for SrTiO$_3$ films grown with fixed Ti flux and with Sr cell temperature of (a) 463 and (c) 462 °C. The substrate temperature was 700 °C. Panel (b) shows RHEED patterns taken along [100] azimuth and [110] azimuth for the surface of (a). The streaks arising from surface reconstruction are indicated by black lines.

**Bias-dependent nanoline images:**

We observe that the STM images of the nanolines are bias-dependent. For positive sample bias voltages greater than about 2.7 V, the images of the nanolines are relatively independent of voltage, as shown in Fig. S3(a), consistent with the reports of Castell and co-workers [5,6]. However, for smaller voltages, the contrast (or height) of these nanolines is found to depend strongly on the sample voltage. When the same surface area is imaged at voltages of +2.1 V or below, then the nanolines become invisible, as shown in Fig. S3(b) and S3(c). This voltage-dependence of the nanoline contrast (height) is found to vary somewhat from sample to sample, e.g. in some samples the nanolines are still visible at +1.5 V but with reduced corrugation height. Nevertheless, we always observe a trend in which the contrast is reduced when sample bias voltage is reduced to values near the CB edge, indicating that the electronic structure of the nanolines at energies near the CB edge differs from that of the surrounding surface.

Tunneling spectra were acquired on the lines, and off the nanolines at the nearby flat surface areas. Off the nanolines, an in-gap peak located near the CB edge is clearly observed, as shown in the left panel of Fig. S3(d). These in-gap states are associated with a transition level of polaronic state of the surface oxygen vacancies, which transitions from a 2+ charged state to + state with one polaron [7]. However, spectroscopic results obtained *on* the nanolines display a significantly less intense in-gap peak, as shown in the right panel of Fig. S3(d).

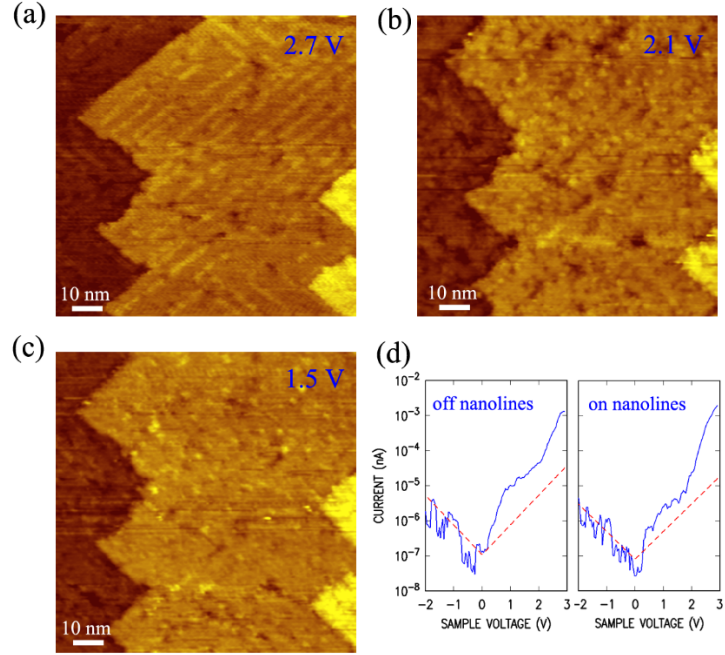

FIG. S3. STM images illustrating a dependence on bias of the nanoline contrast, together with averaged current vs. voltage spectra, for homoepitaxial surfaces of SrTiO$_3$(001) grown with very oxygen deficient conditions (Growth temperature of 800˚C and oxygen partial pressure of $1 \times 10^{-6}$ Torr). The images were acquired from the same surface area, using a tunneling current of 0.03 nA and sample bias voltages of (a) 2.7, (b) 2.1, and (c) 1.5 V. Image size is $100 \times 100$ nm$^2$ for all the three images. The corresponding height-range are 9, 9, 11 Å, respectively. (d) Spectra obtained from a flat, pristine surface area ("off" spectrum) and "on" a nanoline, respectively.

When the sample bias is reduced, the nanolines disappear due to this difference in electronic structure. In other words, the tip-sample separation will be smaller when the tip is positioned on top of the nanolines compared to when it is on top of the pristine surface area at 1.5-2.1 V, in a constant-current imaging mode, since there are fewer state available above the nanoline. Therefore, the apparent height of the nanolines at these biases will be less than its actual height, and hence as the tip moves from the pristine surface area onto the nanolines it will appear as if the nanolines are not present (or only weakly present). Imaging with a higher sample bias, on the other hand, will not produce this height reduction of the nanolines since there is no significant difference in local density of states between the pristine surface area and the nanolines.